# Partitioning and Self-organization of Distributed Generation in Large Distribution Networks

Badr Al Faiya, Stephen McArthur, Ivana Kockar
Department of Electronic and Electrical Engineering
University of Strathclyde
Glasgow, UK
{badr.alfaiya, s.mcarthur, ivana.kockar}@strath.ac.uk

*Abstract*— **Distribution networks will experience more installations of distributed generation (DG) that is unpredictable and stochastic in nature. Greater distributed control and intelligence will allow challenges such as voltage control to be handled effectively. The partitioning of power networks into smaller clusters provides a method to split the control problem into manageable sub-problems. This paper presents a community detection-based partitioning technique for distribution networks considering local DGs, allowing them to be grouped and controlled in a distributed manner by using local signals and measurements. This method also allows each community to control the voltage using only neighboring DGs, and for each community to self-organize to reflect varying DG conditions and to maintain stable control. Simulations demonstrate that the partitioning of the large distribution network is effective, and each community is able to self-organize and to regulate the voltage independently using only its local DGs.**

*Index Terms*-- **Distributed generation, distributed voltage control, community detection, multi-agent systems, self-organization.**

## I. Introduction

The growing penetration of distributed and renewable generation sources in smart grids introduces various technical challenges within distribution networks, such as power quality and voltage control. These challenges can be met using decentralized and distributed control methods based on local information while still maintaining system-level coordination [1]. In this paper, we present a technique for the partitioning of large distribution networks, considering local distributed generation (DG), to enable distributed voltage regulation while dealing with the network time-varying conditions.

In power networks, local elements, including loads, DG, and the network itself, have varying degrees of coupling and influence to each other. In addition, the increasing number of controllable devices, such as DG, the corresponding control variables, and the increasing volume of exchanged information and data in future smart distribution networks are leading to unprecedented complexity in the required control paradigm [2], [3]. The clustering of large power networks into smaller partitions allows splitting the control problem into manageable sub-problems. There is significant prior art in this area. In [4], [5], the K-means algorithm based on electrical distance is used to divide power networks into partitions. The community detection algorithm-based method is used to find optimal partitioning in feeders using reactive power balance degree [6], electrical distance and overvoltage regulation capability [7], and coupling strength [8]. An optimal voltage regulation approach was presented in [9] based on the application of the epsilon-decomposition method to the sensitivity matrix. In [10], a distributed optimization technique for solving the optimal power flow (OPF) problem uses a spectral-clustering algorithm.

Multi-agent system (MAS) based control methods have been used to effectively implement distributed control schemes based on DG. In [11], an agent-based control scheme was introduced by adjusting the output power of the DGs to reach the balance between the demand and supply while providing stability of the voltage and frequency. An MAS-based voltage support model was presented in [12] using DGs in a distribution feeder. Recently, some researchers have attempted to explore MAS and self-organization for power networks. Self-organizing techniques have attracted interest in addressing the uncertainties and dynamic requirements in distributed and complex systems [13]. In addition to its well-known benefit of adaptability, self-organization also has key features such as dynamic and decentralized properties. In [14], a self-organizing communication architecture using an MAS was presented to mitigate cyber-threats on control schemes in smart grids. The energy market problem based on an MAS was investigated in [15] using price and performance management for microgrids. All of these partitioning and distributed control methods are able to tackle network challenges; however, none of them provided grouping of distribution networks by considering the unpredictable DG conditions and the highly dynamic behavior of the emerging smart grids.

To address these challenges, this paper presents a community detection-based partitioning technique that considers the degree of coupling between DGs and nodes in large distribution networks. The voltage control problem is realized by means of a distributed control mechanism based on an MAS architecture and is extended by implementing a dynamic structure within each community to resolve voltage issues using only neighboring DGs, thereby reducing the size of the problem and the interaction requirements within each community. This control scheme also enables each community to self-organize based on network conditions and changes (e.g. DG trips or lost communication). The desired control scheme is

This work was supported by the University of Strathclyde, UK, and Jazan University, Saudi Arabia.

robust to anomalies and uses local interactions to adjust its structure without stopping the system. In addition, self-organization enables decentralization, as only local interactions are allowed. The effectiveness of the algorithm is simulated on a heavily-meshed distribution network with DGs, demonstrating its robustness and autonomy under time-varying DG conditions.

## II. DG-Based Partitioning of Distribution Networks

### A. Index of Partition Quality

Newman's fast algorithm based on a modularity index is one of the most popular and widely used community detection algorithms for large complex networks [16], [17]. The modularity index is a function that gives an indication of the quality of the network's partitioning into communities, expressed as

$$M = \frac{1}{2n} \sum_{i,j} (A_{ij} - k_i k_j / 2n) \delta_{ij} \quad (1)$$

where $n$ is the number of edges; $A_{ij}$ is the edge weight between $i$ and $j$; $k_i$ ($k_j$) is the degree of node $i$ ($j$); and $\delta_{ij}$ equals 1 if nodes $i$ and $j$ are in the same community, otherwise equals 0.

The value $M$ is calculated after each step of joining two communities. The larger the value of the modularity index, the stronger the partition structure. The maximum value of $M$ gives the best partition of the network.

### B. Sensitivity Matrix and DG Based Partitioning

In this study, the modularity index-based algorithm is used to divide a large power network into smaller communities. The sensitivity matrix of the distribution network is used to represent the edge weight information about the degree of coupling between the nodes, represented as

$$\begin{pmatrix} \Delta\theta \\ \Delta V \end{pmatrix} = \begin{pmatrix} \Lambda_{\theta P} & \Lambda_{\theta Q} \\ \Lambda_{VP} & \Lambda_{VQ} \end{pmatrix} \begin{pmatrix} \Delta P \\ \Delta Q \end{pmatrix} \quad (2)$$

with the sensitivity matrix $\Lambda$ written as

$$\Lambda = \begin{pmatrix} \Lambda_{\theta P} & \Lambda_{\theta Q} \\ \Lambda_{VP} & \Lambda_{VQ} \end{pmatrix} \quad (3)$$

To evaluate the strength of a community structure based on DGs, we use the sensitivity matrix to consider the coupling strength of the nodes with nearest DGs. As an example of the partitioning, let us consider the sensitivity submatrix $\Lambda_{VQ}$ in (3) to find the elements of a DG adjacency matrix $D_{VQ}$ that describe the edges that have the highest weight connecting a node $i$ with the nearest DG $j$, which can be expressed as

$$D_{VQ}^{ij} = \begin{cases} 1, & \Lambda_{VQ}^{ij} = \max\{\Lambda_{VQ}^{i1}, \dots, \Lambda_{VQ}^{iN}\}, \\ 0, & \text{otherise,} \end{cases} \quad (4)$$

where $D_{VQ}^{ij}$ represents the element of the DG adjacency matrix between node $j$ and DG $j$, $N$ is the number of DGs in the network, and $\Lambda_{VQ}^{ij}$ is the element of the sensitivity matrix. Thus, the matrix $A_{ij}$ of the modularity index in (1) is implemented using the sensitivity matrix and DG adjacency matrix as

$$A_{VQ}^{ij} = \Lambda_{VQ}^{ij} + D_{VQ}^{ij}. \quad (5)$$

This modular measure of community quality considers the network topology and the degree of coupling between DGs and nodes, enabling each partition to control the voltage independently using its local DGs. The purpose of the partitioning is to reduce the number of variables and constraints within each small community to create a small optimization problem with minimal interaction requirements. The community structure with high modularity has stronger intra-community connections between its nodes. The process and steps to implement the partitioning algorithm are as follows:

*Step 1*: Begin with each node as a separate community.

*Step 2*: For each step, iteratively merge communities in pairs and calculate community quality $M$ using (1) and (5), choosing at each step the community pair that results in the highest $M$ until all communities are merged.

*Step 3*: Finally, the partitioning that results in a local peak in the $M$ value indicates satisfactory partitioning.

The partitioning algorithm is first implemented to partition the distribution network into communities. After determining the communities, each community maintains the voltage levels in its community within certain limits using its local DG, as discussed below.

### C. Distributed Voltage Control within Each Community

In this study, we are interested in controlling the voltage in each community using local DGs. After determining the communities as described above for a large network, some communities may still contain a large number of DGs that can have varying influence on community nodes. Moreover, it is possible that not all DGs in a community have strong couplings with all intra-community nodes. Thus, to find which DGs are most influential to a node, we define the "neighboring DG" to the nodes within a community for which the voltages are strongly affected by the output of the DG. We analyze each community using only the updated community DG adjacency matrix to describe DG that have the greatest influence on nodes. To enable each nodal voltage in the community to be controlled by the neighboring DGs that have the highest influence on the node, we define the subsets of neighboring DG using the community DG matrix $D_{com}$ in (1) as follows

$$A_{com}^{ij} = D_{com}^{ij} \quad (6)$$

where

$$D_{com}^{ij} = \begin{cases} 1, & \Lambda_{com}^{ij} = \max\{\Lambda_{com}^{i1}, \dots, \Lambda_{com}^{iN}\}, \\ 0, & \text{otherise.} \end{cases} \quad (7)$$

This enables each community to maintain the voltage levels within certain limits by calculating the optimal adjustments of only the neighboring DGs to the voltage violation. It is noted that each subset contains one or more DGs depending on the size and topology of the community. Based on the sensitivity matrix $\Lambda$, we can calculate adjustments to the DG reactive power, $x_Q = Q_r - Q_0$, or active power, $x_P = P_r - P_0$, to control the voltage at a particular node from an initial voltage $V_0$ to a reference voltage $V_r$ as

$$V_r = V_0 + A_{VP} \cdot x_P + A_{VQ} \cdot x_Q \quad (8)$$

where $Q_0$ and $Q_r$ are the reactive power outputs before and after voltage regulation, respectively, and $P_0$ and $P_r$ are the DG active power outputs before and after voltage regulation, respectively.

## III. Deployment of the MAS Architecture

### A. Types of Agents

After implementing the partitioning method on the distribution network as described in Section II-B, agents are partitioned into small communities, with the communication links between agents being determined autonomously. In order to analyze the partitioning and distributed control algorithm, and given the space and scope of this paper, it is assumed that the communities are fixed in this study. However, the partitioning functionality allows dynamic re-partitioning of communities in the future. To deploy this proposed approach, the system contains three types of agents: community agent (CA), bus agent (BA), and distributed generation agent (DA). After determining the communities of the network, each community is assigned one CA that acts as its control agent. Each agent uses its behavior and knowledge to autonomously manage and coordinate its activities with only those agents involved while maintaining a stable state of the system, as described below.

#### 1) Community Agent (CA):

The CA finds solutions for the community voltage control problem using the DGs that are closer to the voltage violation. First, the CA defines the subsets of nodes and neighboring DGs by using (1) and (7) to describe how the community nodes are affected by the output of the community DGs. In addition, the CA updates the elements of the community DG matrix $D_{com}^{ij}$ based on its knowledge or when triggered by other agents or changes in the network. For example, this can be triggered by a DA when its DG has been tripped. This updates the community DG matrix in accordance with (7). The agents self-organize in new subsets to adapt to the network and DG conditions without stopping or restarting the MAS platform and its agents.

When a CA receives a voltage violation message from a BA in its community, it determines the optimal adjustments to the generation of the neighboring DGs using a linear programming (LP) algorithm that is integrated within each CA. Thus, the CA has knowledge of: a) the DAs and the surplus capacity of each DG in the same community; b) the voltage sensitivity submatrix to define how each DG influences the voltages of the nodes; c) the sensitivity coefficients, obtained from (3), that determine how a DG affects the voltage of the primary and secondary sides of the network transformers; and d) the voltage upper and lower normal operating levels (in this paper, between 0.95 pu and 1.05 pu). The third constraint is to avoid the tripping of the network protectors that are used to prevent reverse active power flows through the network from the secondary side to the primary side, as shown in (13) and (15).

For the reactive power and voltage control problem using $\Lambda_{VQ}$, the community can control the voltage optimally by minimally decreasing or increasing the reactive power outputs of the DGs. For the LP problem, the objective function is [9]:

$$\text{Max: } Min\{x_i\} \text{ (to control overvoltage)} \qquad (9)$$
$$\text{Min: } Max\{x_i\} \text{ (to control undervoltage)} \qquad (10)$$

subject to the following constraints:

$$\begin{cases} V_l \leq V_0 + \Lambda_{VQ} \cdot x \leq V_u \\ x \leq Q_{Sur} \\ 0 \leq \theta_{p0} + \Lambda_{\theta_pQ} \cdot x - (\theta_{s0} + \theta_{shift} + \Lambda_{\theta_sQ} \cdot x) \end{cases} \qquad (11)$$

For the active power and voltage control problem using $\Lambda_{VP}$, the CA determines the optimal adjustments to the generation of the neighboring DG to control the voltage using the following objective function:

$$\text{Max: } Min\{x_i\} \qquad (12)$$

s.t.

$$\begin{cases} V_l \leq V_0 + \Lambda_{VP} \cdot x \leq V_u \\ x \leq P_{Sur} \\ 0 \leq \theta_{p0} + \Lambda_{\theta_pP} \cdot x - (\theta_{s0} + \theta_{shift} + \Lambda_{\theta_sP} \cdot x) \end{cases} \qquad (13)$$

To solve this LP problem in the same way as the standard LP problem, we can add a slack variable $y$ to bring it into the same form as the standard LP problem. Let us consider overvoltage control in the PFC mode as an example:

$$\text{Max: } y \text{ (to control overvoltage)} \qquad (14)$$

s.t.

$$\begin{cases} V_l \leq V_0 + \Lambda_{VQ} \cdot x \leq V_u \\ x \leq Q_{Sur} \\ 0 \leq \theta_{p0} + \Lambda_{\theta_pQ} \cdot x - (\theta_{s0} + \theta_{shift} + \Lambda_{\theta_sQ} \cdot x) \\ x_i \geq y \; (i = 1 \sim n, n \text{ neighboring DAs}) \end{cases} \qquad (15)$$

In the above equations, $x$ is the vector of all $x_i$; $x_i$ is the power generation adjustment of the $i$-th DG agent; $V_u$ and $V_l$ are the upper and lower voltage limits, respectively; $Q_{Sur}$ and $P_{Sur}$ are the surplus capacities of the DG; $\theta_{p0}$ and $\theta_{s0}$ are the initial values of the network transformer voltage angles on the primary side and secondary side, respectively; $\Lambda_{\theta_pQ}$ and $\Lambda_{\theta_sQ}$ are the sensitivity submatrices of the reactive power adjustments and the transformer voltage angles on the primary and secondary sides, respectively; and $\Lambda_{\theta_pP}$ and $\Lambda_{\theta_sP}$ are the sensitivity submatrices of the active power adjustments and the transformer voltage angles on the primary and secondary sides, respectively.

When the CA solves the LP problem in its community, only the DGs with high influence on the voltage violation node participate in voltage control, further reducing both the size of the optimization problem and the interaction requirements. After determining how to resolve the voltage violation problem, the CA communicates the control adjustments to the involved DAs to request an increase or decrease in the DG outputs and to restore the community's voltage to be within the acceptable operating limits. However, if the DG is unable to resolve the voltage issue (e.g., a DG is not available or has lost communication), the CA updates the $D_{com}^{ij}$ elements based on DG conditions. Similarly, if a DG is tripped, the DA informs the CA to update $D_{com}^{ij}$ based on the available DGs. As a result, the community agents reconfigure and self-organize to regulate the voltage.

*2) Bus Agent (BA):*

The BAs monitor the status of buses and have knowledge of the voltage violation constraints, as specified by the voltage upper and lower acceptable operating levels (in this paper, between 0.95 pu and 1.05 pu). When a BA observes a voltage violation at its bus, it requests its CA to resolve the community voltage issue.

*3) Distributed Generation Agent (DA):*

Each DA represents its DG that is connected to the distribution network and performs control actions to adjust its generation. The DA receives adjustment signals from its CA to maintain acceptable voltage limits. The DA has knowledge of its DG constraints, such as generation capacity, and shares its availability and constraints with other agents. For instance, if a DG is not available (e.g., is disconnected from the network), the DA informs the CA so that the CA can maintain stable control through the self-organizing mechanism.

### B. Implementation of Dynamic and Distributed Voltage Regulation

After partitioning the network into communities, the agents in each community coordinate to realize and maintain distributed control, as follows.

*Step 1):* Each CA initializes its community DG matrix and finds how DGs influence nodes by employing (1) and (7).

*Step 2):* Each BA begins monitoring its bus for voltage violations; if such a violation occurs, the BA sends a violation message to its CA.

*Step 3):* The CA checks the violation messages sent by its community and determines output adjustments as expressed in (9)–(15) for the related DAs. Then, the CA sends the control actions to the DAs to optimize its output to bring the voltage in that community back to within normal limits.

*Step 4):* If a DG is not able to regulate the voltage (e.g., a DG has been tripped or has reached its available surplus capacity), the CA updates the community DG adjacency matrix based on the available DGs. Subsequently, the agents in the community regroup and self-organize in the newly identified subsets to control the voltage. When the network returns to normal operation (e.g., the DG is back online), the agents reorganize and return to Step 2 to continue monitoring and controlling the community.

## IV. APPLICATION TO A LARGE DISTRIBUTION NETWORK

### A. Self-Organizing MAS Implementation

The MAS architecture was implemented using the Simulation of Agent Societies 2 (Presage2) framework [18], which offers agent communication capabilities and improved autonomy. Presage2 provides the flexibility to design self-organizing MASs able to meet the requirements of electrical network control and management applications.

### B. The Test System

The system under study is a model of a real heavily-meshed secondary network, used from [9], containing 2083 nodes, 224 network transformers (13.8 kV to 480 V or 216 V), and 311 PQ loads. As shown in Fig. 1, the primary feeders are at 13.8 kV and contain 1043 nodes, while the secondary network contains the remaining 1040 nodes at 216 V or 480 V. In this network, the 311 DGs are installed at the load buses in the secondary network.

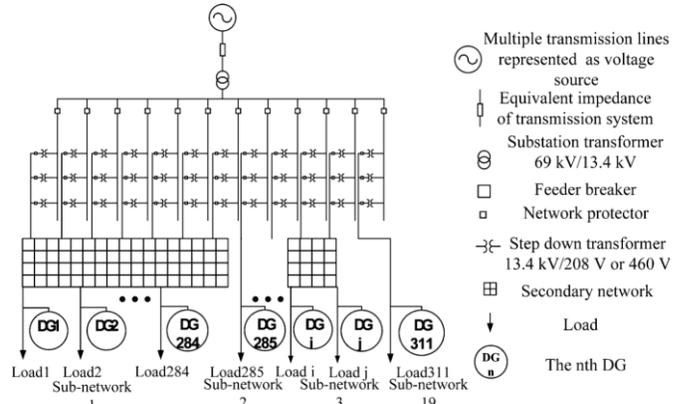

Fig.1. The structure of the test system [9].

## V. RESULTS AND DISCUSSIONS

In the following, we perform the proposed partitioning and distributed control algorithm in the test system to validate the autonomy and adaptability of the system.

### A. Partitioning of The Model System

The technique is tested in the network to demonstrate effective partitioning and reactive power-voltage control. Table I shows that the network is divided into 56 communities. Thus, the agents in the secondary network are grouped into 56 communities, with one CA is assigned in each community. The CA informs the BAs and DAs so that each agent is linked with agents in the same community. The table also show that communities have varying number of DGs and nodes.

TABLE I
RESULTS OF COMMUNITY NUMBER AND SIZE

| Community No. | No. of Nodes | No. of DGs | Community No. | No. of Nodes | No. of DGs |
|---|---|---|---|---|---|
| 1 | 35 | 5 | 29 | 43 | 29 |
| 2 | 16 | 5 | 30 | 31 | 15 |
| 3 | 15 | 3 | 31 | 27 | 14 |
| 4 | 18 | 5 | 32 | 6 | 2 |
| 5 | 19 | 7 | 33 | 15 | 4 |
| 6 | 22 | 4 | 34 | 13 | 2 |
| 7 | 20 | 8 | 35 | 14 | 2 |
| 8 | 17 | 6 | 36 | 28 | 6 |
| 9 | 16 | 5 | 37 | 9 | 2 |
| 10 | 16 | 3 | 38 | 29 | 8 |
| 11 | 21 | 8 | 39 | 25 | 6 |
| 12 | 13 | 3 | 40 | 9 | 4 |
| 13 | 10 | 2 | 41 | 11 | 4 |
| 14 | 23 | 3 | 42 | 33 | 12 |
| 15 | 18 | 6 | 43 | 17 | 7 |
| 16 | 35 | 6 | 44 | 21 | 4 |
| 17 | 17 | 4 | 45 | 9 | 4 |
| 18 | 28 | 9 | 46 | 6 | 2 |
| 19 | 11 | 4 | 47 | 29 | 11 |
| 20 | 34 | 5 | 48 | 22 | 8 |
| 21 | 13 | 1 | 49 | 15 | 3 |
| **22** | **25** | **8** | 50 | 10 | 2 |
| 23 | 29 | 12 | 51 | 24 | 5 |
| 24 | 23 | 3 | 52 | 11 | 3 |
| 25 | 29 | 10 | 53 | 13 | 3 |
| 26 | 8 | 2 | 54 | 7 | 2 |
| 27 | 13 | 5 | 55 | 7 | 2 |
| 28 | 5 | 2 | 56 | 7 | 1 |

TABLE II
SUBSETS OF COMMUNITY 23 BEFORE DG TRIP

| Subset No. | List of agents |
|---|---|
| 1 | 232 233 234 235 236 237 238 249 |
| 2 | 239 240 241 242 384 947 |
| 3 | 243 244 245 246 |
| 4 | 247 248 364 365 366 727 949 |
| 5 | 385 386 387 388 |

TABLE III
SUBSETS OF COMMUNITY 23 AFTER DG TRIP

| Subset No. | List of agents |
|---|---|
| 1 | 232 233 234 235 236 237 238 249 |
| 2 | 239 240 241 242 384 385 386 387 388 947 |
| 3 | 243 244 245 246 |
| 4 | 247 248 364 365 366 727 949 |

*B. Self-organizing Mechanism for Distributed Voltage Regulation*

Taking community number 23 as an example as shown in Table II, the community determines DGs and nodes within the community for which the DGs have more influence on those nodes. To demonstrate self-organization and distributed voltage regulation method, we assume a disconnection of DG unit 386, in subset 5, which causes a voltage violation to appear in the community. As a result, the DA 386 informs its CA 23 about the trip event so that the CA can find the updated $D_{com}$ matrix. As shown in Table III, the CA now identifies 4 subsets. The agents then self-organize and continue monitoring and regulating the voltage.

After the DG trip, a voltage violation is detected by BA 386 which informs the corresponding control agent CA 23. Thus, CA 23 finds a solution by running the LP algorithm using the subset DGs and sends the optimal adjustment messages to DAs 239, 340, and 341 to inject reactive powers of 0.24 pu, 0.22, and 0.24 pu, respectively. Therefore, out of the 12 DG agents in this community, only three DGs are involved in the control problem. This further reduces the size of the LP control problem from 12 DGs to only 3 DGs, while simultaneously reducing the interactions required between only the involved agents. The voltage profile before and after the control in shown on Fig 5.

## VI. CONCLUSIONS

This paper presents a partitioning and self-organizing distributed voltage control method for large distribution networks with DGs. The system is implemented using an MAS framework to control the voltage using only DGs neighboring the voltage violation through local interactions of agents in a cooperative way. Each community cluster can also adapt to time-varying DG conditions to maintain stable control without re-engineering the complete system. Simulation results validated the effectiveness of the presented technique based on a heavily-meshed distribution network to control the voltage by each community independently using only the local DGs.

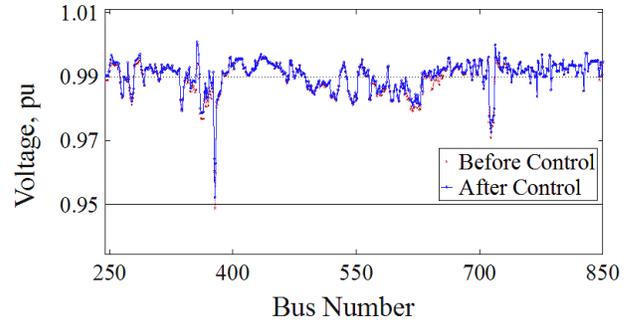

Fig. 2. Voltage profiles of busses showing violation at bus 385 and regulation by the neighboring DGs.